\documentclass[%
 reprint,
superscriptaddress,
 amsmath,amssymb,
 aps,
floatfix,
]{revtex4-2}

\usepackage{graphicx}% Include figure files
\usepackage{dcolumn}% Align table columns on decimal point
\usepackage{bm}% bold math
\begin{document}

\preprint{APS/123-QED}

\title{Anderson localization and reentrant delocalization of tensorial elastic waves in two-dimensional fractured media}

\author{Qinghua Lei}
 \email{qinghua.lei@erdw.ethz.ch}
 \affiliation{Department of Earth Sciences, ETH Zurich, Zurich, Switzerland}

\author{Didier Sornette}
 \affiliation{Department of Earth Sciences, ETH Zurich, Zurich, Switzerland}
 \affiliation{Department of Management, Technology and Economics, ETH Zurich, Zurich, Switzerland}
 \affiliation{Department of Physics, ETH Zurich, Zurich, Switzerland}
 \affiliation{Institute of Risk Analysis, Prediction and Management, Academy for Advanced Interdisciplinary Studies, Southern University of Science and Technology, Shenzhen, China}

\date{\today}% It is always \today, today,
             %  but any date may be explicitly specified

\begin{abstract}
We study two-dimensional tensorial elastic wave transport in densely fractured media and document transitions from propagation to diffusion and to localization/delocalization. For large fracture stiffness, waves are propagative at the scale of the system. For small stiffness, multiple scattering prevails, such that waves are diffusive in disconnected fracture networks, and localized in connected ones with a strong multifractality of the intensity field. A reentrant delocalization is found in well-connected networks due to energy leakage via evanescent waves and cascades of mode conversion.
\end{abstract}
\maketitle

Understanding elastic wave transport in crustal rocks is crucial for earthquake prediction and geophysical exploration \cite{stein2003}. Fractures, as an important type of crustal heterogeneities \cite{sornette2006,lei2016}, can strongly affect the wavefield behavior \cite{pyraknolte1990,vlastos2003,hamzehpour2016}. When a wave packet enters an inhomogeneous system, it can exhibit various regimes such as propagation, diffusion, and localization \cite{ishimaru1978,sheng2006,sornette1989a}. The wave is propagative at scales less than the mean free path length, beyond which it becomes diffusive if the disorder of the medium is not too strong \cite{sornette1989b}. If the disorder is strong enough so that the Ioffe-Regel criterion \cite{ioffe1960} is met, the wavefield enters a regime without effective transport, known as Anderson localization \cite{anderson1958}, as a result of very strong interferences between multiply scattered waves by the disordered medium acting as effective random mirror cavities \cite{sornette1989c,rossum1999}. This transition was originally predicted for electron wave functions to explain the metal-insulator transition in disordered crystals, but was later realized as a generic phenomenon for all wave types \cite{anderson1985,lagendijk2009}. Anderson localization has been experimentally demonstrated for microwaves \cite{dalichaouch1991}, light waves \cite{sperling2013,schwartz2007}, acoustic waves \cite{hu2008,sornette1989d}, and matter waves \cite{billy2003}. It is speculated that Anderson localization of elastic waves may also happen in the Earth{'}s crust \cite{lagendijk2009}, as implied by the field evidence of weak localization \cite{larose2004} and coda localization \cite{friedrich2005,aki2000}. However, the conditions under which Anderson localization occurs in fractured crustal rocks remain unknown.

In this Letter, we study the transport and localization of tensorial elastic waves in fractured media based on two-dimensional (2D) numerical simulations. We focus on the heterogeneity effect induced by fractures and limit our scope to pure elastic problems. We capture complex wave phenomena in fracture systems arising collectively from the transmission, reflection, refraction, and diffraction by numerous individual fractures. We show a full spectrum of wave energy transport transitioning from propagation to diffusion and to localization/delocalization.

\textit{Model approach and setup.}\textemdash The numerical model uses the finite element method to solve the 2D time-domain elastodynamic equation that governs tensorial waves in linear elastic solids under a plane-strain condition \cite{lei2020}. Fractures are represented as discrete linear elastic interfaces of vanishing thickness and non-welded contact based on the displacement discontinuity method \cite{schoenberg1980}, such that across each fracture, stress is continuous but displacement is discontinuous controlled by the fracture stiffness \cite{pyraknolte1990}. More details of our model including the governing equations are in the Supplemental Material \cite{supplemental_material}.

We generate synthetic fracture networks embedded in an otherwise isotropic and homogeneous matrix within a domain $\Omega$ of size $2L\times2L$. The matrix has a density $\rho$, and P and S wave velocities $V$ and $V^{\prime}$, respectively, with $V/V^{\prime} = 1.73$. The location and orientation of fractures are purely random, whereas their length is constant as $2l = L/5$. The fracture network connectivity is quantified by the percolation parameter $p = Nl^2/L^2$ \cite{balberg1984}, where $N$ is the number of fractures. Fracture system transforms from a disconnected to a connected state as $p$ exceeds the percolation threshold $p_\mathrm{c} \approx 6$ \cite{balberg1984}. We consider three $p$ values, $2$, $6$, and $10$, corresponding to a non-connected, critically-connected, and well-connected network, respectively. The network of a higher $p$ is constructed by adding fractures into that of a lower $p$. For each $p$ value, 100 realizations are generated for ensemble averaging.

The fracture networks are placed in a cell surrounded by absorbing layers \cite{pettit2014} that suppress unwanted reflections at the model boundary so as to mimic an unbounded space. We define a Cartesian coordinate system with the origin at the domain center and two axes (x and y) orthogonal to the boundary. A point source is located at the origin to excite purely circular P waves by applying a force-type sinusoid signal with the wavelength $\lambda = L/10$. The displacement amplitude of the incident wave at the source is $A_0$ ($A_0/L \to 0$). We consider two excitation scenarios using a one-cycle wave pulse or a continuous wave train. A wide range of fracture stiffness $k$ is explored with the normalized stiffness $\tilde{k} = k/(\omega Z) = 0.001$, $0.01$, $0.1$, $1$, $10$, and $100$, where $\omega = 2\pi V/\lambda$ is the angular frequency and $Z = \rho V$ is the seismic impedance. The normal and shear stiffness components are identical. We normalize the time $t$ as $\tilde{t} = (t-t_0)/(L/V-t_0)$, where $t_0 = \lambda/(2V)$ is the half-period of the wave emitted from the source. We compute up to $\tilde{t}  = 1$ and $10$ for the one-cycle and continuous excitation simulations, respectively. More details of our model including material properties, numerical settings, a sensitivity analysis of spatial/temporal discretization, and a demonstration of the statistical convergence are in the Supplemental Material \cite{supplemental_material}.

\begin{figure*}
\centering
\includegraphics[scale=0.38]{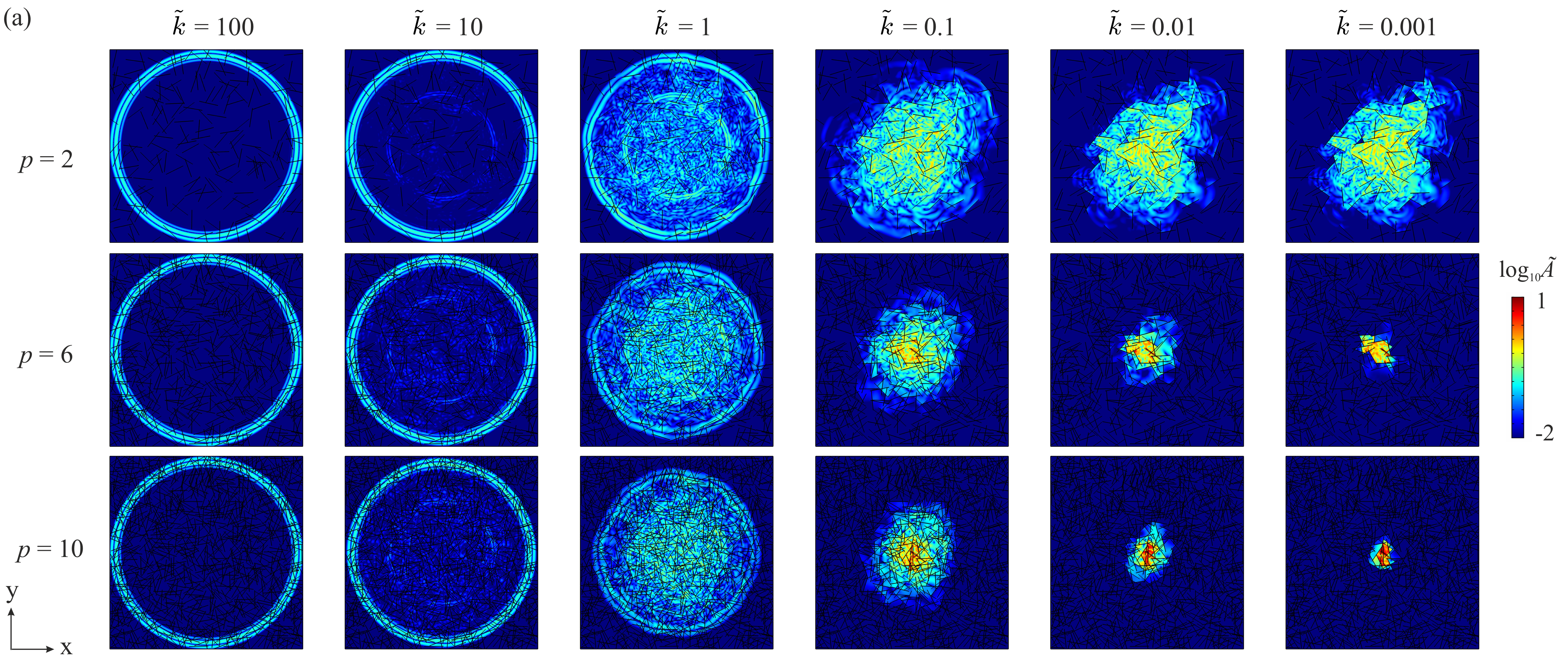}
\includegraphics[scale=0.62,trim={0cm 0cm 0.3cm 0cm},clip]{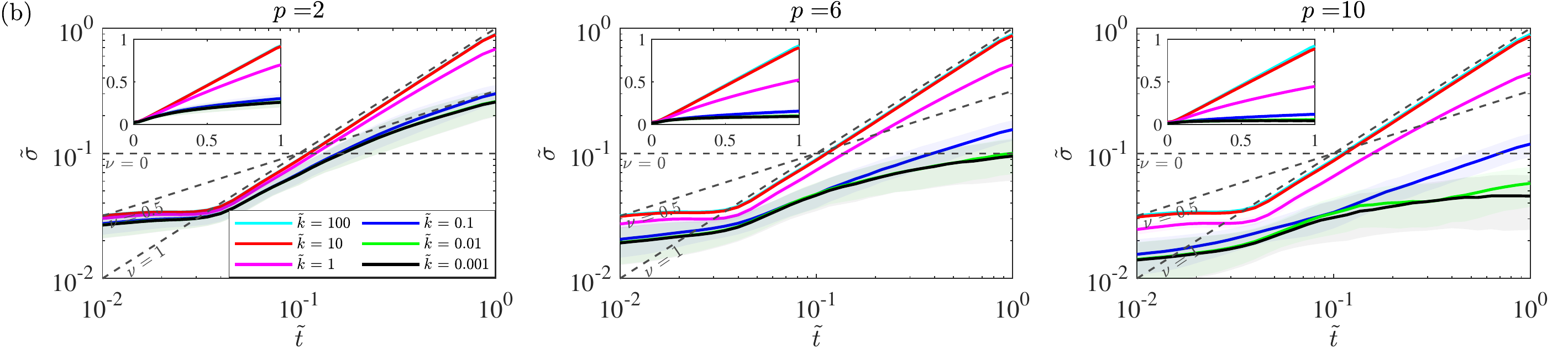}
\caption{\label{fig:fig1}(a) Wavefields illustrated by the normalized amplitude $\tilde{A}$ at time $\tilde{t} = 1$ after exciting a circular, one-cycle P wave from the center of fractured media with different percolation parameters $p$ and normalized fracture stiffnesses $\tilde{k}$. (b) Temporal variation of the normalized gyradius $\tilde{\sigma}$; shaded area indicates the standard deviation over 100 realizations; insets show $\tilde{\sigma}$ versus $\tilde{t}$ in linear scales.}
\end{figure*}

\textit{Excitation of a one-cycle wave.}\textemdash Fig.~\ref{fig:fig1}a shows the wavefield spatial distribution by plotting the normalized amplitude $\tilde{A}$ (local amplitude $A$ normalized by $A_0$) at $\tilde{t}$ = 1, for which no wave has escaped out of the domain. The wavefield evolution is quantified by the normalized gyradius \cite{sperling2013,sebbah1993} (Fig.~\ref{fig:fig1}b), $\tilde{\sigma} = \frac{1}{L}[\iint_\Omega \tilde{I} r^2 dx dy / \iint_\Omega \tilde{I} dx dy]^{1/2}$, where $\tilde{I} = \tilde{A}^2$ is the normalized wave intensity, and $r$ is the distance between the wave position ($x$, $y$) and the source (0, 0). The temporal scaling of $\tilde{\sigma}$ is expected to obey $\tilde{\sigma} \sim \tilde{t}^\nu$ with $\nu$ taking $1$, $0.5$, and $0$ for ballistic propagation, normal diffusion, and Anderson localization, respectively.

\begin{figure}
\centering
\includegraphics[scale=0.48]{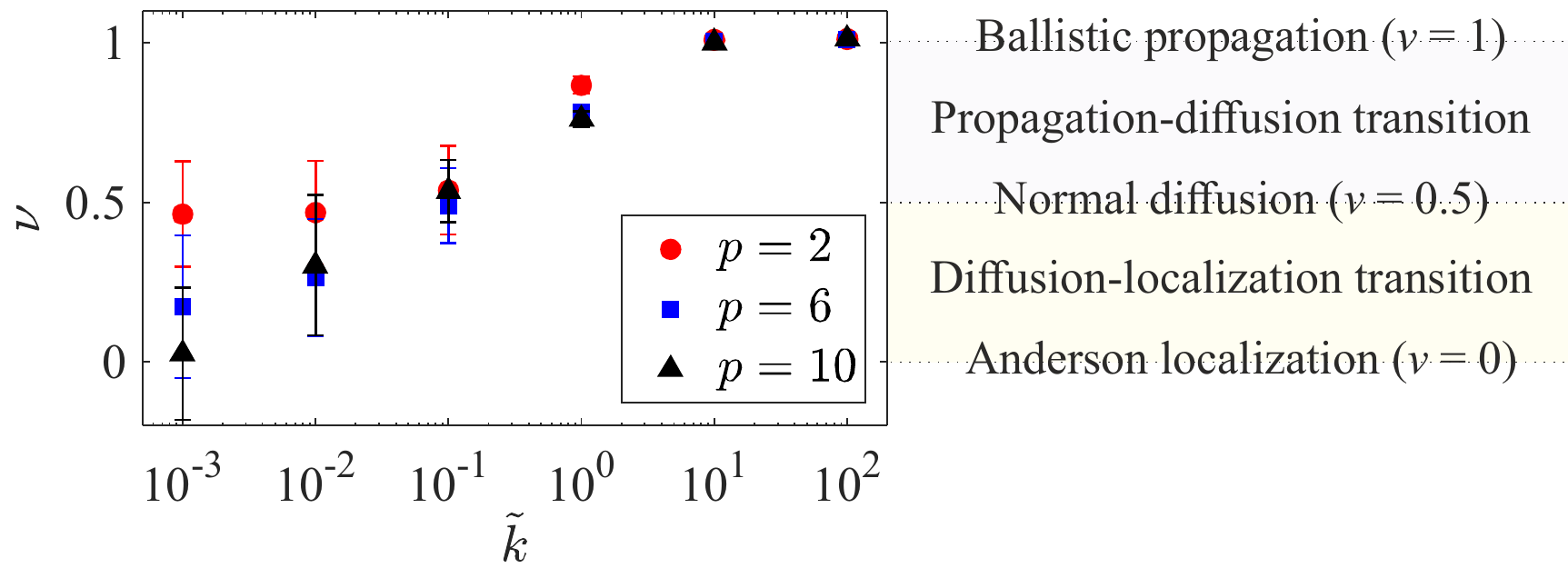}
\caption{\label{fig:fig2}Dependence of the exponent $\nu$ as a function of the normalized fracture stiffness $\tilde{k}$ for fractured media with different percolation parameters $p$. Markers and error bars indicate the mean and standard deviation over 100 realizations.}
\end{figure}

For $\tilde{k} = 100$, fractures have no visible influence on the wavefield, whose circular wavefront is fully preserved irrespective of $p$. For $\tilde{k} = 10$, the wavefront still keeps a circular shape, although some scattered waves are left behind it, which are more noticeable as $p$ increases. For both $\tilde{k} = 100$ and $10$, $\tilde{\sigma}$ scales linearly with $\tilde{t}$ ($\nu = 1$); some deviation from linearity exists in the early stage ($\tilde{t}<0.05$) since the source excitation has not completed its full-period yet. For $\tilde{k} = 1$, the wavefront becomes considerably distorted with extensive scattered waves emerging in its interior, such that the wavefield is characterized by the fast propagation of frontal waves and slow diffusion of coda waves. Increasing $p$ tends to cause a reduction of the wavefront speed and an enhancement of coda generation. Consequently, $\tilde{\sigma}$ scales sublinearly with $\tilde{t}$ ($\nu < 1$). For $\tilde{k} = 0.1$, the wavefront is essentially destroyed and significant wave energy is backscattered/backreflected, phenomena which become more pronounced as $p$ increases; the scaling of $\tilde{\sigma}$ with $\tilde{t}$ is compatible with $\nu = 0.5$ (diffusive transport dominates). For $\tilde{k} = 0.01$, waves can still escape from the disconnected network of $p = 2$ through the gaps between fractures, but are trapped near the source by the networks of $p = 6$ and $10$; wave diffusion becomes anomalous ($\nu < 0.5$), indicating the emergence of weak localization \cite{sebbah1993}\textemdash a precursor of Anderson localization \cite{sheng2006,sornette1989c,rossum1999}. For $\tilde{k} = 0.001$, Anderson localization occurs in the well-connected network ($p = 10$), qualified by a saturation of $\tilde{\sigma}$ ($\nu \to 0$) at the late stage ($\tilde{t}>0.5$). Nevertheless, waves in the less connected networks ($p = 2$ and $6$) still exhibit abnormal diffusion ($0 < \nu < 0.5$). The control of fracture stiffness on the wavefield transition from propagative to diffusive and to localized modes is demonstrated by the monotonic decrease of $\nu$ with the reduction of $\tilde{k}$ (Fig.~\ref{fig:fig2}).

\begin{figure*}
\centering
\includegraphics[scale=0.375]{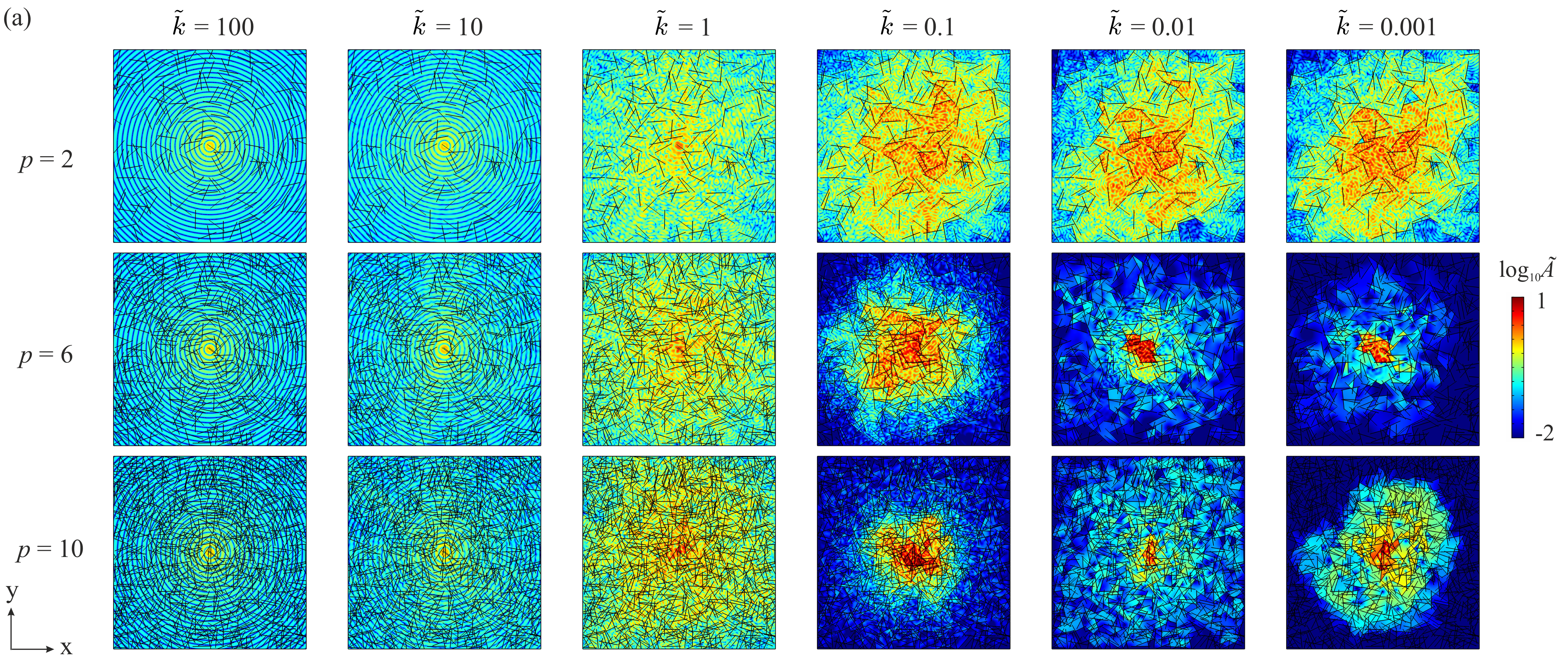}
\includegraphics[scale=0.61,trim={0cm 0cm 0.3cm 0cm},clip]{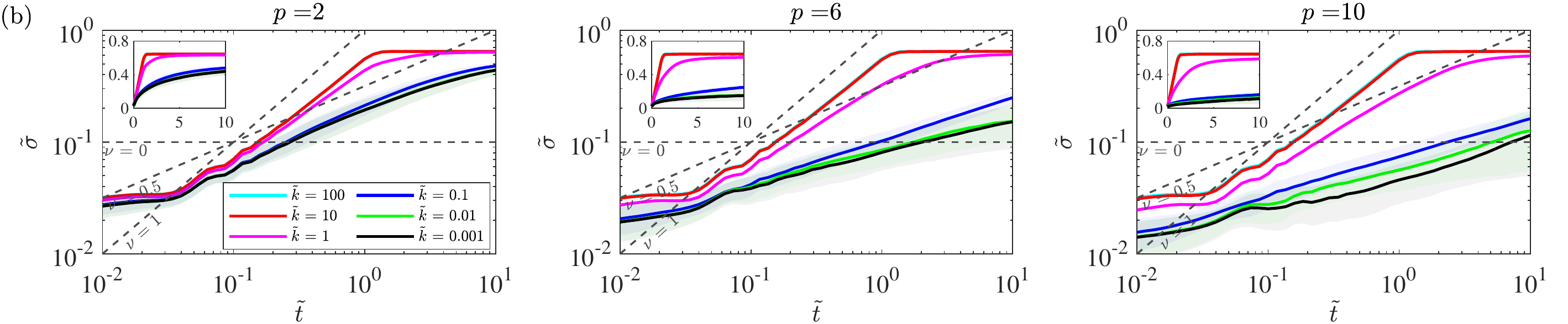}
\caption{\label{fig:fig3}(a) Wavefields illustrated by the normalized amplitude $\tilde{A}$ at time $\tilde{t} = 10$ after exciting a circular, continuous P wave from the center of fractured media with different percolation parameters $p$ and normalized fracture stiffnesses $\tilde{k}$. (b) Temporal variation of the normalized gyradius $\tilde{\sigma}$; shaded area indicates the standard deviation over 100 realizations; insets show $\tilde{\sigma}$ versus $\tilde{t}$ in linear scales.}
\end{figure*}

\textit{Excitation of a continuous wave.}\textemdash Fig.~\ref{fig:fig3}a shows the spatial distribution of $\tilde{A}$ at $\tilde{t} = 10$, while Fig.~\ref{fig:fig3}b plots the temporal evolution of $\tilde{\sigma}$. For $\tilde{k} = 100$ and $10$, the wavefield exhibits a ripple-like pattern consisting of a sustained wave train radially propagating outwards; $\tilde{\sigma}$ scales linearly with $\tilde{t}$ for $\tilde{t} \leq 1$ ($\nu \approx 1$), after which a plateau is reached because those waves extending beyond the finite-sized domain are absorbed and the wavefield converges to a quasi-steady state. For $\tilde{k} = 1$, the wavefield becomes scattered, more noticeable as $p$ increases. However, the spatial footprints of waves are still relatively homogeneous due to the dominant propagating part of the wave energy ($0.5 < \nu < 1$). For $\tilde{k} = 0.1$, multiple scattering prevails, resulting in highly-diffusive wavefields ($\nu \approx 0.5$); furthermore, the diffusion slows down with an increased $p$. For $\tilde{k} = 0.01$ and $0.001$, the wavefield in the network of $p = 2$ is similar to that for $\tilde{k} = 0.1$ ($\nu \approx 0.5$), but that in the network of $p = 6$ becomes more subdiffusive or weakly localized ($\nu < 0.5$). In contrast, the wavefield of $p = 10$ becomes more localized around the source while also globally more diffusive with $\nu \approx 0.5$. Such a reentrant phenomenon of delocalization is attributed to the dominance of evanescent waves \cite{stein2003}, which leak wave energy out of the closed “cavity” formed by interconnected fractures. Since the properties of the matrix at the two sides of a fracture are the same, in order to generate evanescent waves, the only possible mechanism is that a S wave impinges on the fracture at an incidence angle exceeding the critical angle $\theta_c = \arcsin(V^\prime/V) = 35.3^{\circ}$, associated with a decreased transmission of converted P waves \cite{pyraknolte1990}. This prerequisite can be easily found when strong multiple scattering occurs, promoting an equipartition of P and S waves \cite{hennino2001} (although the source only excites P waves) and distributing them into arbitrary directions \cite{ishimaru1978}. The generated evanescent waves then propagate along fractures as interface waves \cite{pyraknolte1987,pyraknolte1992} at a speed asymptotically approaching that of Rayleigh waves for low $\tilde{k}$.

\begin{figure}
\centering
\includegraphics[scale=0.3]{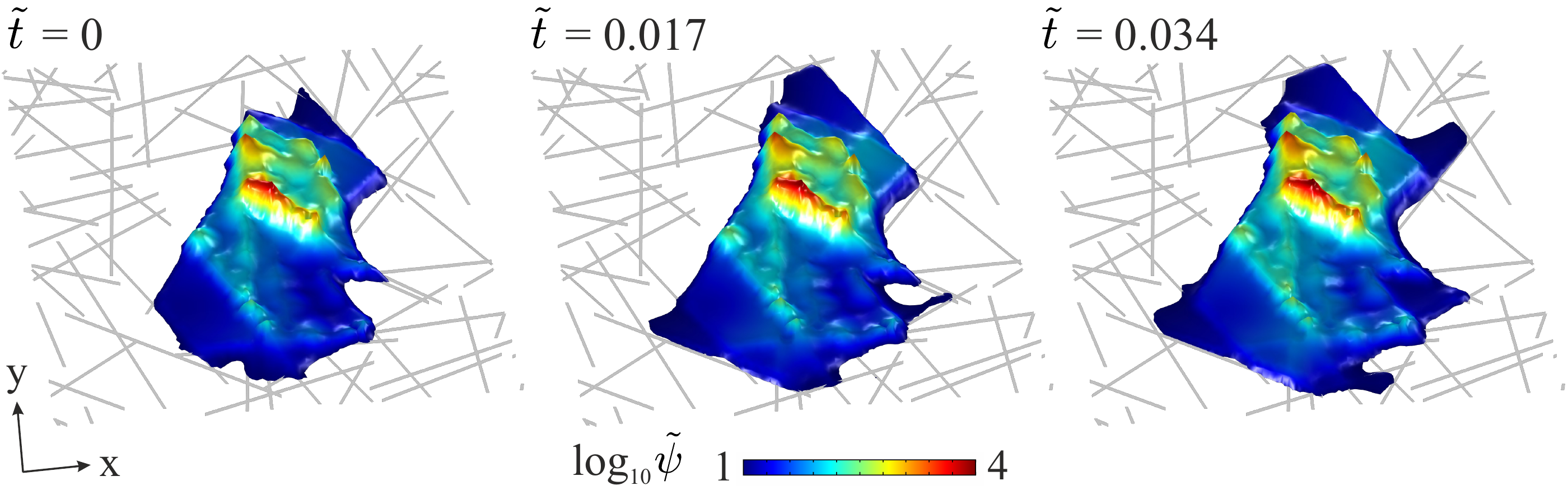}
\caption{\label{fig:fig4}Transmitted wave energy $\tilde{\psi}$ at different times $\tilde{t}$ in the near-field ($-0.2 \leq x/L \leq 0.2$ and $-0.2 \leq y/L \leq 0.2$) of the source, which excites a continuous P wave into the fracture network with the percolation parameter $p$ = 10 and normalized fracture stiffness $\tilde{k} = 0.001$.}
\end{figure}

To further elucidate the emergence and consequence of evanescent waves, we calculate the transmitted wave energy at any given point within $\Omega$ as $\tilde{\psi} = \frac{V}{L} {\int_{0}^{t} \tilde{I} dt}$, which is the normalized temporal integration of the wave intensity. Fig.~\ref{fig:fig4} shows the spatial distribution of $\tilde{\psi}$ at different times $\tilde{t}$ in a well-connected network of $p = 10$ and $\tilde{k} = 0.001$. The wavefield is highly trapped in the vicinity of the source by a locally-connected subnetwork ($\tilde{t} = 0$). However, some waves (evanescent waves) manage to escape the cavity and travel along fractures as interface waves, marked by a fingering pattern in the $\tilde{\psi}$ field ($\tilde{t} = 0.017$). These interface waves are then scattered at fracture intersections with some of the energy converted to P and S waves depending on the intersection angle, similar to Rayleigh wave scattering at corners \cite{bremaecker1958} (the only difference is that fracture interface waves involve weakly-coupled fracture-walls \cite{pyraknolte1987,pyraknolte1992} instead of free-surfaces). The converted P and S wave energy is then accumulated and trapped in a new cavity ($\tilde{t} = 0.034$), until a secondary leakage occurs, again via evanescent waves. This cascading process driven by iterative mode conversions eventually leads to a progressive leakage of wave energy out of closed cavities formed by interconnected fractures of low stiffness. Delocalization is suppressed if the network is less connected because of reduced number of fracture intersections \cite{balberg1984} and enlarged size of cavities (the strength of evanescent waves decays exponentially away from the interface \cite{stein2003} and thus the energy cannot efficiently accumulate if the cavity is much larger than the wavelength). Delocalization is not conspicuous in the one-cycle excitation scenario, since the cascading leakage requires sufficient energy supply to overcome multiple layers of closed cavities and sufficient time to develop multiple scattering and equipartition.

\begin{figure}
\centering
\includegraphics[scale=0.56]{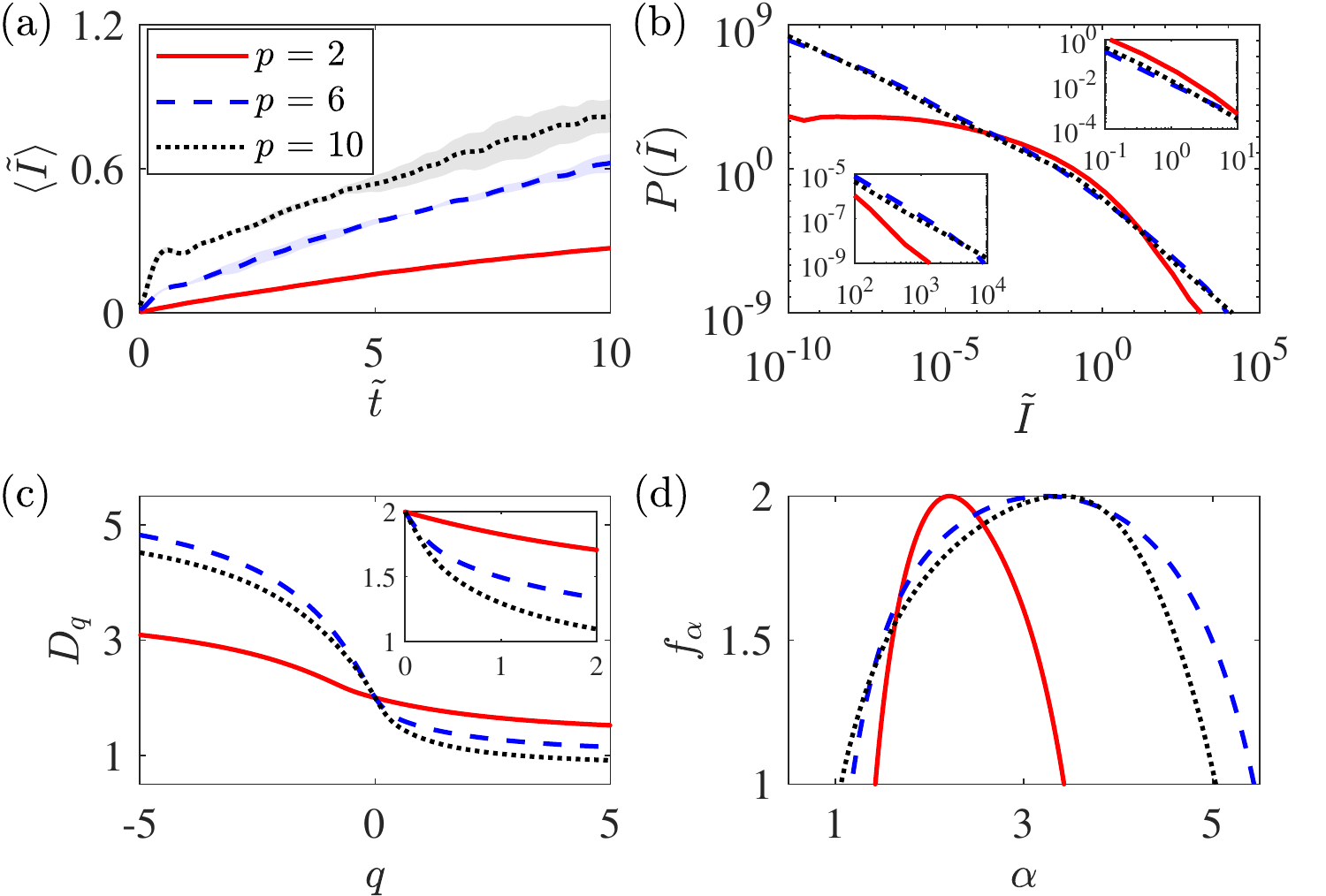}
\caption{\label{fig:fig5}(a) Spatially-averaged wave intensity $\langle \tilde{I} \rangle$ versus time $\tilde{t}$; (b) probability density function of normalized intensity $P(\tilde{I})$; (c) generalized fractal dimension $D_q$ of order $q$ and (d) singularity spectrum $f_\alpha$ versus the H\"{o}lder exponent $\alpha$ for the near-field intensity at $\tilde{t} = 10$. Insets give zoomed-in views.}
\end{figure}

Finally, we investigate the distribution of wave intensity in fracture networks with $\tilde{k} = 0.001$. Fig.~\ref{fig:fig5}a shows the temporal variation of the spatially-averaged intensity $\langle \tilde{I} \rangle$ of the entire wavefield. The energy grows continuously in the connected networks of $p = 6$ and $10$ due to wave localization, while that of $p = 2$ tends to saturate because of outward diffusion. The probability density function $P(\tilde{I})$ (Fig.~\ref{fig:fig5}b) shows that, as $p$ increases, the intensity distribution changes from a lognormal-like to a power law-like behavior. $P(\tilde{I})$ for $p = 10$ has a marginally heavier tail than that for $p = 6$ (bottom left inset), suggesting a slightly stronger localization in the former system. A higher proportion of intermediate intensity ($0.1 \leq \tilde{I} < 1$) is noticed for $p = 10$ compared to $p = 6$ (upper right inset) due to wave delocalization.

We also perform a multifractal analysis to characterize the spatial heterogeneity and hierarchy of the wavefield in the near-field of the source ($-0.2 \leq x/L \leq 0.2$ and $-0.2 \leq y/L \leq 0.2$), since multifractality of wave energy is a hallmark of Anderson localization \cite{castellani1986,schreiber1991,evers2008}. The generalized fractal dimension $D_q$ of order $q$ (Fig.~\ref{fig:fig5}c) is defined from the moment $M_q = \sum_{i=1}^{n} \Theta_i^q \sim \varepsilon^{(q-1)D_q}$ \cite{sornette2006}, where the spatial domain is partitioned into $n$ boxes of size $\varepsilon$, $\Theta$ is the sum of local intensities within the $i$th box normalized by the total intensity. The multifractal spectrum \cite{sornette2006} (Fig.~\ref{fig:fig5}d) is determined by computing the H\"{o}lder exponent $\alpha = \sum_{i=1}^{n} \Theta_i^q \ln \Theta_i /(M_q \ln \varepsilon)$, which is related to the singularity spectrum $f_\alpha$ via the Legendre transform $f_{\alpha} = \alpha q - D_q(q-1)$. As $p$ increases from 2 to 6 and then to 10, the high energy content of the waves becomes more and more localized, as revealed by the enhanced nonlinearity of $D_q$ for $q>0$ (Fig.~\ref{fig:fig5}c) and the increased width of $f_\alpha$ in the small $\alpha$ region (Fig.~\ref{fig:fig5}d). The low-to-intermediate energy content also becomes more localized when $p$ increases from 2 to 6, associated with a steeper $D_q$ for $q<0$ (Fig.~\ref{fig:fig5}c) and a wider $f_\alpha$ for large $\alpha$ (Fig.~\ref{fig:fig5}d). However, as $p$ further increases from $6$ to $10$, $D_q$ for $q < 0$ flattens and $f_\alpha$ at large $\alpha$ narrows, implying that low-to-intermediate energy content is more diffuse and homogeneously partitioned over space in the well-connected network, due to wave delocalization.

To conclude, we have shown transitions of elastic wave transport in fractured media from propagation to diffusion and to localization/delocalization, governed by fracture stiffness and network connectivity. Waves propagate ballistically when the stiffness is large. As the stiffness decreases, multiple scattering occurs, leading to diffusive transport in disconnected fracture networks and localization in connected ones. We have documented a reentrant delocalization in well-connected networks due to evanescent waves and cascades of mode conversion. We have demonstrated that wave localization is qualified by a saturated growth of wavefield gyradius and a strong multifractality of intensity field, whereas delocalization destroys these characteristics. This research opens the door to understanding complex wave phenomena in fractured media and has important implications for many geophysical problems.

\bibliography{prl_manuscript}% Produces the bibliography via BibTeX.

\begin{thebibliography}{38}
\bibitem [{sup()}]{supplemental_material}%
  \BibitemOpen
  \href@noop {} {\bibinfo {title} {See {S}upplemental {M}aterial at [{URL} will
  be inserted by publisher]}},\ \bibinfo {note} {{f}or more details about the
  governing equations, model setup, model validation, sensitivity analysis, and
  statistical convergence}\BibitemShut {NoStop}%
\end{thebibliography}

\end{document}